\documentclass{aa}     
\usepackage{txfonts}
\usepackage{graphicx}
\usepackage{amssymb}
\usepackage[T1]{fontenc}

\usepackage{natbib}

\def\ds {{$\delta$~Scuti~}}
\def\dss {{$\delta$~Scuti~stars}}

\def\bceph {{$\beta$~Cephei}}

\def\teff {{T_{\mathrm{eff}}}}
\def\vsini {{\vr\!\sin\!i}}
\def\vr {{v}}

\def\robar {{\bar \rho}}
\def\robarsol {{\robar_{\sun}}}
\def\Oi {{\Omega_{\mathrm{i}}}}

\def\kms {{\mathrm{km}\,\mathrm{s}^{-1}}}
\def\msol {{\mathrm{M}_\odot}}

\def\ratio {{\Pi_{1/0}}}
\def\ratiorot {{\Pi_{1/0}\,(\Omega)}}
\def\pu {{\Pi_1}}
\def\po {{\Pi_0}}

\begin{document}

   \title{The role of rotation on Petersen Diagrams. The $\ratiorot$ period ratios}

   \titlerunning{The role of rotation on Petersen Diagrams}
   \authorrunning{Su\'arez, Garrido \& Goupil}

   \author{J.C. Su\'arez\inst{1,2}
   \thanks{Associate researcher at institute (2), with finanantial 
	      support from Spanish <<Consejería de Innovación, Ciencia 
	      y Empresa>> from the <<Junta de Andalucía>> local government.}
   \and R. Garrido\inst{1}
   \and M.J. Goupil\inst{2}}

   \offprints{J.C Su\'arez~\email{jcsuarez@iaa.es}}

   \institute{Instituto de Astrof\'{\i}sica de Andaluc\'{\i}a (CSIC), CP3004, Granada, Spain 
	      \and 
	      Observatoire de Paris, LESIA, UMR 8109, Meudon, France}

   \date{Received ... / Accepted ...}

   \abstract{The present work explores the theoretical effects of rotation in
             calculating the period ratios of double-mode radial pulsating
	     stars with special emphasis on high-amplitude \dss\ (HADS). 
	     Diagrams showing these period ratios vs. periods of 
	     the fundamental radial mode have been employed as a good 
	     tracer of non-solar metallicities and are known as Petersen diagrams (PD).
	     In this paper we consider the effect of moderate rotation on both 
	     evolutionary models and oscillation frequencies and we show that
	     such effects cannot be completely neglected as it has been done
	     until now. In particular it is found that even for low-to-moderate rotational 
	     velocities (15--$50\,\kms$), differences in period ratios of some hundredths
	     can be found. The main consequence is therefore the confusion 
	     scenario generated when trying to fit the metallicity of a given star
             using this diagram without a previous knowledge of its rotational
	     velocity.
            }

\maketitle


\section{Introduction\label{sec:intro}}
Radial pulsators, and more particularly double mode pulsators, have been
extensively studied using the well known Petersen diagrams (from now on called
PD). Such diagrams show the ratio between the fundamental radial mode 
and the first overtone as a function of the fundamental mode. 
Typically, the stars concerned are double-mode Cepheids, RR Lyrae
and high-amplitude \dss\ (HADS).
Recently, the analysis of data from large-scale projects like OGLE \citep[Optical
Gravitational Lensing Experiment][]{OGLE_Szy,OGLE_Udal}, 
NSVS \citep[Northern Sky Variability Survey][]{NSVS04},
ASAS \citep[All Sky Automated Survey][]{ASAS02,ASAS03} or 
MACHO \citep{MACHO00_hads} has permitted deeper studies of
the observational properties of such double-mode pulsators.

These period ratios were firstly studied by \citet{Petersen73,Petersen78}, 
and they have been used
for decades as metallicity indicators, as a function of stellar mass and age,
 to test mass-luminosity and/or radius-luminosity relations. In addition,
 they also have been used, for instance, to determine the distance 
 modulus to the SMC \citep{Kovacs00SMC_a}.

\begin{figure}
   \hspace{-0.50cm}
   \includegraphics[width=8.8cm]{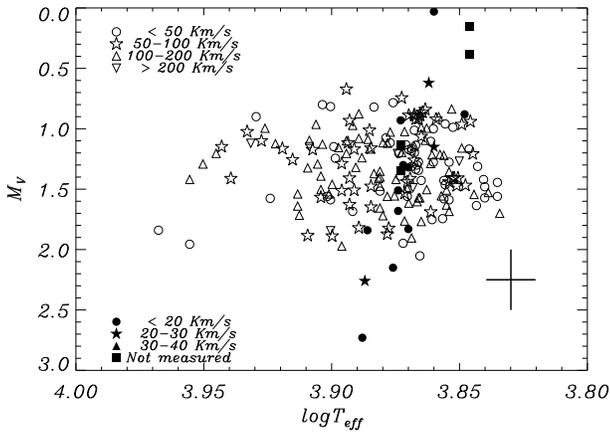}
   \caption{Absolute magnitudes $M_{\mathrm{V}}$ as a function of the effective temperature
            of all known \dss\ (empty symbols), compared with those of all double-mode
	    HADS known up to date (filled symbols). Different symbol types represent
	    the observed $\vsini$ ranges \citep[taken from][]{Rodriguez00}.
	    The cross represents typical errors on absolute magnitudes and
	    effective temperature for \dss.}
   \label{fig:vsiniHadsDSC}
\end{figure}

While non-radial pulsators like \bceph\ or low-amplitude \dss\ (LADS) 
generally show moderate to fast rotational velocities ($\vsini$), the
double-mode radial pulsators can be considered as slow-to-moderate
rotating stars. 
This fact has lead to neglect
systematically the effect of rotation on theoretical period-ratios,
in particular F-1O ratios in PD. However, we remind the reader that
such objects are generally faint enough as to make difficult the
observation of their rotational velocities ($\vsini$). 
Figure~\ref{fig:vsiniHadsDSC} shows the location of all
known double-mode HADS \citep[from][]{Poretti05hads}
in the HR diagram, compared with the location of all LADS with 
measured $\vsini$. As can be seen, the narrow band (in effective
temperature ) occupied by 
double-mode HADS overlaps the region where LADS are located.
In luminosity, HADS occupy a broader range. Unfortunately, 
the number of known HADS constitutes a very poor sample, 
specially when compared with LADS. This means that we cannot 
discard, a priori, the possibility of HADS in a wider range of
effective temperatures and present larger $\vsini$ measurements.
In addition to this, although most of HADS present 
$\vsini\leq 20\,\kms$, this represent a lower limit of their
rotational velocities. When varying the angle of inclination
of the star $i$, velocities up to $50\,\kms$ could be reached.


From the theoretical side, only a few works examined partially 
the effect of rotation
on period ratios and mainly focused on very rapidly rotators.
In \citet{PH95_pd},
second order effects of rotation on
oscillation periods were taken into account in order to discriminate
between radial and non-radial modes. The authors expressed the relative
period change as 
$\delta\Pi_{n}^{\mathrm{(rot)}}/\Pi_n=Z_n(\Pi_{\mathrm{rot}})/\Pi_n)^{-2}$,
where $\Pi_n$ is the unperturbed period of the mode with radial order 
$n$, and $\Pi_{\mathrm{(rot)}}$ being 
 the rotation period. The coefficient $Z_n$ depends on the radial order of the
mode and on the structure of the star. Such values were estimated interpolating
from computations derived for a polytropic model with index 3.
Recently, \citet{Alosha03} studied the behaviour of period ratios
of radial modes when near degeneracy effects due to rotation are included 
for a typical $1.8\,\msol$ \ds\ stellar model. He showed that very large and 
non-regular perturbations to such ratios are expected to occur.
More recently, \citet{Sua05altairII} proposed a limit of 
validity of the perturbation theory (up to second order) in terms of
rotational velocity for rotating models. Such limit is given by the
behaviour of period ratios when near degeneracy is considered, which clearly
complicates the naive interpretation of the PD. 

In the present work, we aim at analysing the consequences of neglecting
the effect of rotation on radial period ratios even for low rotational
velocities. To do so, up-to-date techniques taking properly into account 
the rotation in the modelling (in equilibrium models and in oscillation frequencies) are used.
Similarly as done in \citet{Alosha03} and \citet{Sua05altairII} near 
degeneracy is also considered.

The paper is structured as follows: A general description of the modelling,
focusing on how rotation is taken into account is given in Sect.~\ref{sec:models}.
Fundamental-to-first harmonic ratios in presence of rotation, $\ratiorot$, are
introduced in Sect.~\ref{sec:pet-diag}, and a discussion of their 
impact on PD analysis is proposed in Sect.~\ref{sec:impact_Z}. 
Finally, the conclusions are given in 
Sect.~\ref{sec:conclusions}.

\section{The modelling\label{sec:models}}

The evolutionary code CESAM \citep{Morel97} is used, which 
is particularly adapted for our purposes. The numerical precision and the mesh grid 
(around 2000 mesh points given in the basis of B-splines) of equilibrium models 
have been adapted according to the oscillation computation requirements. 

Following \citet{KipWeig90}, a first order effect of rotation is taken into account
in equilibrium equations. In particular, the spherical symmetric contribution
of the centrifugal acceleration is included by means of an effective gravity
$g_{\mathrm{eff}}=g-{\cal A}_{c}(r)$, where $g$ represents the local gravity 
component, $r$ the radial distance and ${\cal A}_{c}(r)=\frac{2}{3}\,r^2\,\Omega$,
the centrifugal acceleration of matter elements at a distance $r$ 
from the centre of the star. This spherically symmetric 
contribution of the rotation does not change the 
shape of the hydrostatic equilibrium equation. Although, the non-spheric
components of the centrifugal acceleration are not considered, they are
included as a perturbation in the oscillation computation. 
The total angular momentum of models is assumed to
be globally conserved along the evolution of the star. 
Input physics has been adapted for intermediate mass
stars.
\begin{figure}
   \includegraphics[width=8.8cm]{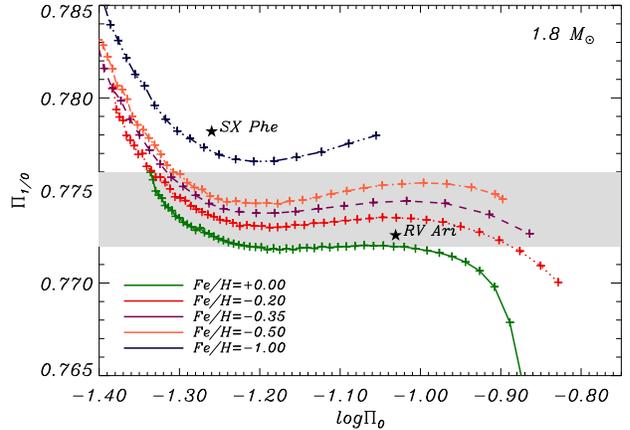}
   \caption{Typical PD ($\po$ in $\mathrm{d}$) containing different tracks of 
            $1.8\,\msol$ evolutionary models computed with different 
	    initial metal content [Fe/H], from solar composition (bottom, solid line) to 
	    -1.00 (top, triple dot-dashed line). Crosses represent non-rotating models.
	    The shaded area corresponds to typical values found for Pop.~I stars.
	    The two filled star symbols represent the observed $\ratio$ of
	    the double-mode high-amplitude \dss\ RV~Ari and SX~Phe, as an illustration
	    of (Pop.~I) and (Pop.~II) stars respectively. (For clarity, colours
	   are used in the \emph{on-line} version of the paper).}
   \label{fig:classic_PD}
\end{figure}

Theoretical oscillation spectra are computed from the equilibrium models described in the previous
section. For this purpose the oscillation code \emph{Filou} \citep{filou,SuaThesis} is used. This
code, based on a perturbative analysis, provides adiabatic oscillations corrected for the effects 
of rotation up to second order (centrifugal and Coriolis forces). 

Furthermore, for moderate--high rotational velocities, the effects of near degeneracy are expected
to be significant \citep{Soufi98}. Two or more modes, close in frequency, are rendered \emph{degenerate} 
by rotation under certain conditions, corresponding to selection rules. In particular these rules select modes
 with
the same azimuthal order $m$ and degrees $\ell$
differing by 2 \citep{Soufi98}. If we consider two generic modes $\alpha_1\equiv(n,\ell,m)$ and
$\alpha_2\equiv(n^\prime,\ell^\prime,m^\prime)$ under the aforementioned conditions, near degeneracy
occurs for $|\sigma_{\alpha_1}-\sigma_{\alpha_2}|\leq\sigma_{\Omega}$,
where $\sigma_{\alpha_1}$ and $\sigma_{\alpha_2}$ represent the eigenfrequency associated to modes
$\alpha_1$ and $\alpha_2$ respectively, and $\sigma_\Omega$ represents the stellar rotational
frequency. In certain cases, such effect may be dominant in the behaviour of the
radial period ratios studied here \citep{Alosha03, Sua05altairII}. However, due to its complexity,
we believe such effects should be analysed separately (Su\'arez et al., in prep.) and,
for the sake of clarity, they have not been included in the present work.

\section{The $\ratiorot$ ratios\label{sec:pet-diag}}

In general, low order radial period ratios can be considered as 
dependent of the distribution of mass (or density) throughout the star, and
the thermodynamical properties of the stellar matter. 
In Fig.~\ref{fig:classic_PD}, a classic PD for $1.8\,\msol$ evolutionary tracks
is depicted. It illustrates the well-known dependence on the metallicity of $\ratio=\pu/\po$
ratios similar as those shown for instance in \citet{Petersen73}, 
\citet{PetDalsgaard96} and \citet{PetDalsgaard96_2}.
As can be seen, the period ratios increase when decreasing the stellar initial metal 
content. This property is commonly used to discriminate, in the context of radial pulsators,
Pop. I from Pop. II stars. The shaded region indicates the typical $\ratio$
values found for Pop. I stars, in the range of $\ratio=[0.772, 0.776]$. As a reference,
the observed period ratios of two stars: RV Ari (Pop. I) and SX~Phe (Pop. II) are
also depicted \citep[values obtained from][ and references therein]{Poretti05hads}.
\begin{figure}
   \includegraphics[width=8.8cm]{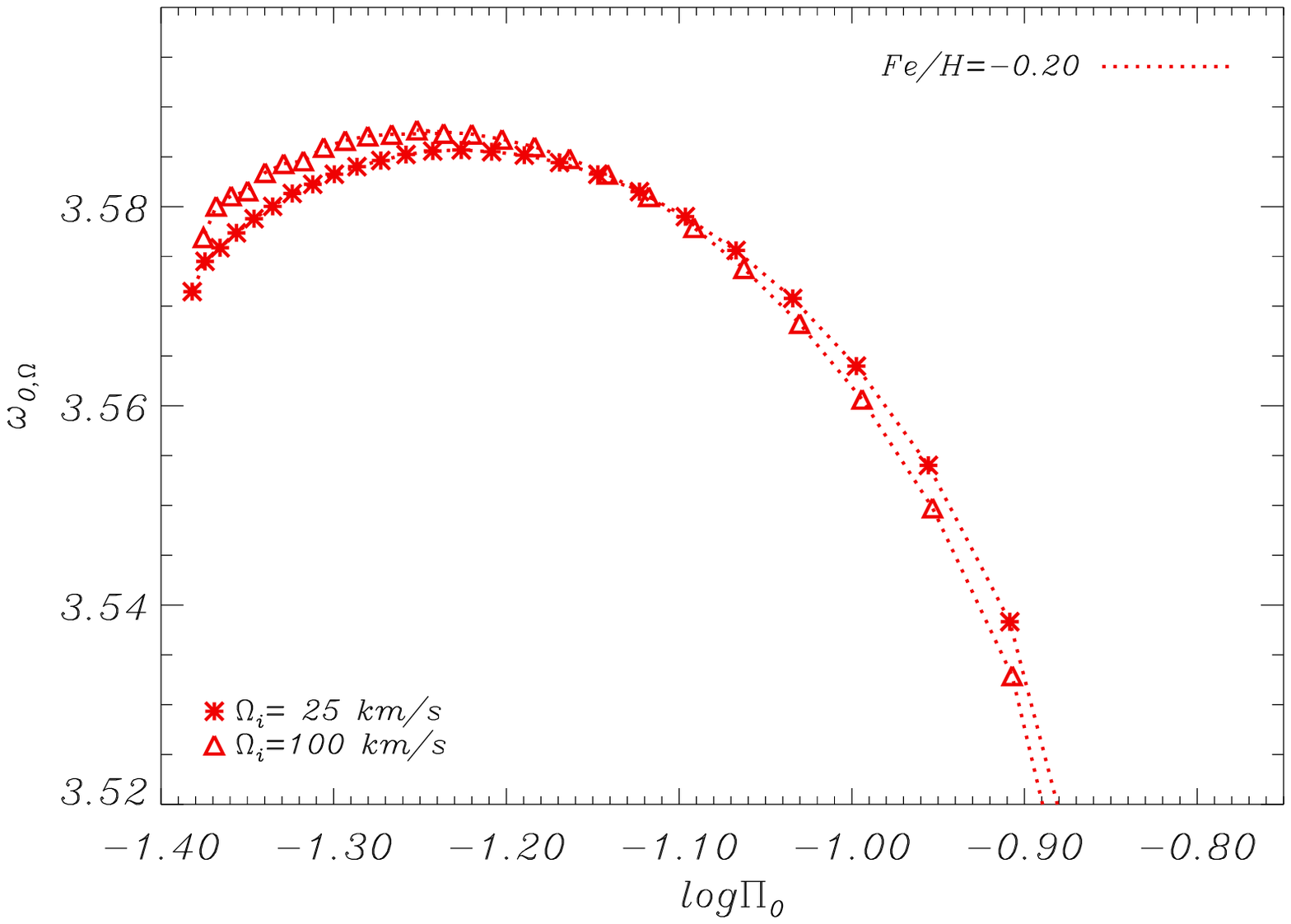}
   \includegraphics[width=8.8cm]{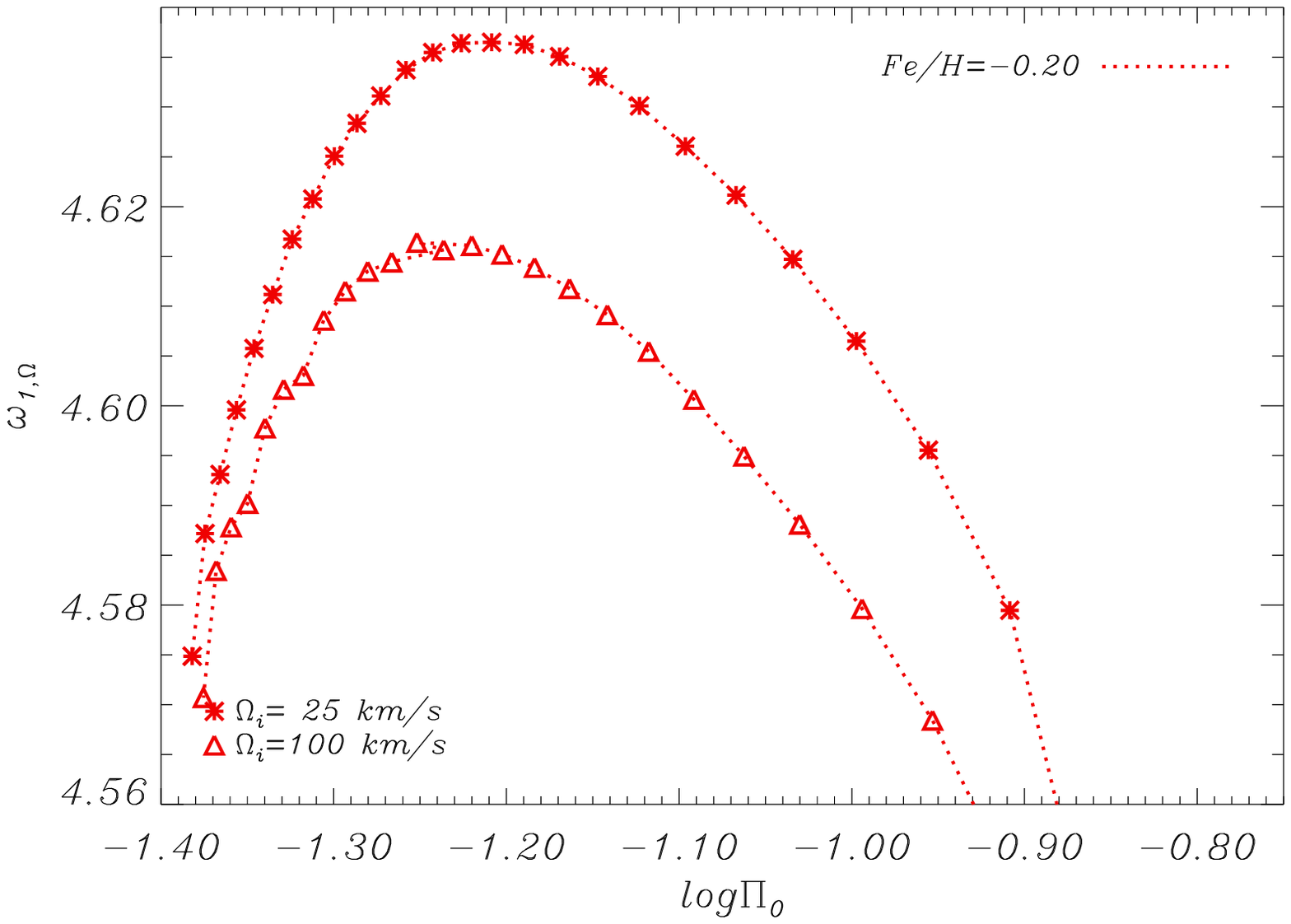}   
   \caption{Fundamental radial order mode $\omega_{0,\Omega}$
           (top panel) and the first overtone $\omega_{1,\Omega}$ 
	   (bottom panel) normalised frequencies
	   as a function of the logarithm of the fundamental radial
	   order period (in $\mathrm{d}$). (For clarity, colours
	   are used in the \emph{on-line} version of the paper).}
   \label{fig:w0w1}
\end{figure}
For a given chemical composition the period ratio $\ratio$ is determined by mass
and the radius (or $\po$ which is scaled by the mean density $\robar$, and then
$$\ratio=\ratio\,(M, R, Z)\,.$$
In addition, period ratios can be easily written in terms of observed quantities through the
so-called pulsation constant $Q_{n,\ell} = \Pi_{n,\ell}\sqrt{\robar/\robarsol}$.
In the context of radial modes, this constant can be expressed as
$$Q_{1,0} = \Pi_{1,0}\,\Big(\frac{g}{g_{\sun}}\Big)^{1/2}\,
                         \Big(\frac{L}{L_{\sun}}\Big)^{-1/4}\,
			 \Big(\frac{\teff}{T_{\mathrm{eff,\sun}}}\Big)\,,
	         \label{eq:def_Qnl}$$
allowing us to assign to each observed radial period $\Pi_{n,0}$ a value $Q_{n,0}$,
which can be compared with theoretical predictions.

However, it is well known that rotation modifies the structure of stars
and thereby the cavity where modes propagate. It is thus plausible to 
consider the period ratios to be $\Omega$-dependent. Moreover, 
the angle of inclination of the star is generally unknown (only rotational 
projected velocities ($\vsini$) are provided by observations), and thereby 
it follows 
$$\ratiorot=\ratio\,(M, R, Z, \Omega(i))\,.$$
From a theoretical point of view, following \citet{Soufi98, Sua05rotcel_sub}, the adiabatic 
oscillation eigenfrequencies $\omega_{n,\ell,m}$ can be expressed in
terms of a perturbative theory as:
$$\omega_{n,0,0}=\omega_{n,0,0}^{(0)}+\omega_{n,0,0}^{(2)}\label{eq:omega}$$
for radial modes ($\ell=0, m=0$), where $\omega_{n,0,0}^{(0)}$ represents the 
unperturbed frequency and
$\omega_{n,0,0}^{(2)}$ the second order correction term. 
Both terms implicitly include the symmetric component of the 
centrifugal force (mainly the departure from sphericity of the star)
given by the equilibrium model.
First order correcting terms $\omega_{1,0,0}^{(1)}$ (Coriolis force effect) 
are proportional to the azimuthal order $m$, and therefore they are
zero for radial modes. For shortness, from now, only the subscript $n$ is
kept. Figure~\ref{fig:w0w1} illustrates the effect of rotation on
normalised\footnote{The normalisation constant is $(G\,M/R^3)^{1/2}$ which is 
proportional to the model mean density $\robar$.} frequencies corresponding to 
the fundamental mode $\omega_{0,\Omega}$ (top panel) and the first overtone 
$\omega_{1,\Omega}$ (bottom panel) for a given metallicity. 
As can be seen, the first overtone frequencies are visibly more affected
by rotation than the fundamental ones. Without entering in details
(a theoretical work analysing the behaviour of $\ratiorot$ is currently
in preparation) this suggests the possibility that
$\omega_{1,\Omega}$ is proportional to the distribution of the density
inside the star (affected by rotation) rather than simply the mean density.
\begin{figure*}
 \begin{center}
   \includegraphics[width=9cm]{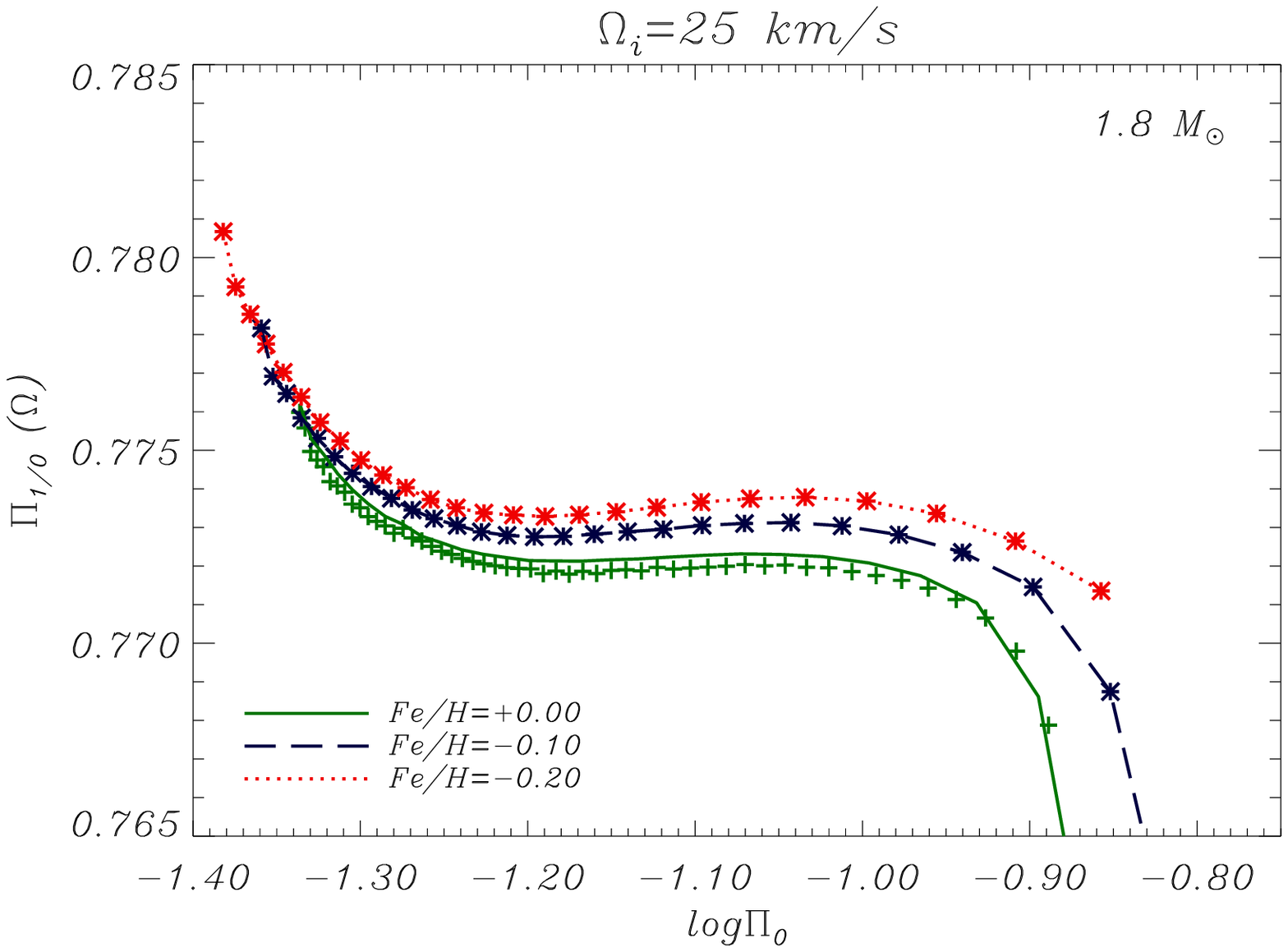}\hspace{-0.40cm}
   \includegraphics[width=9cm]{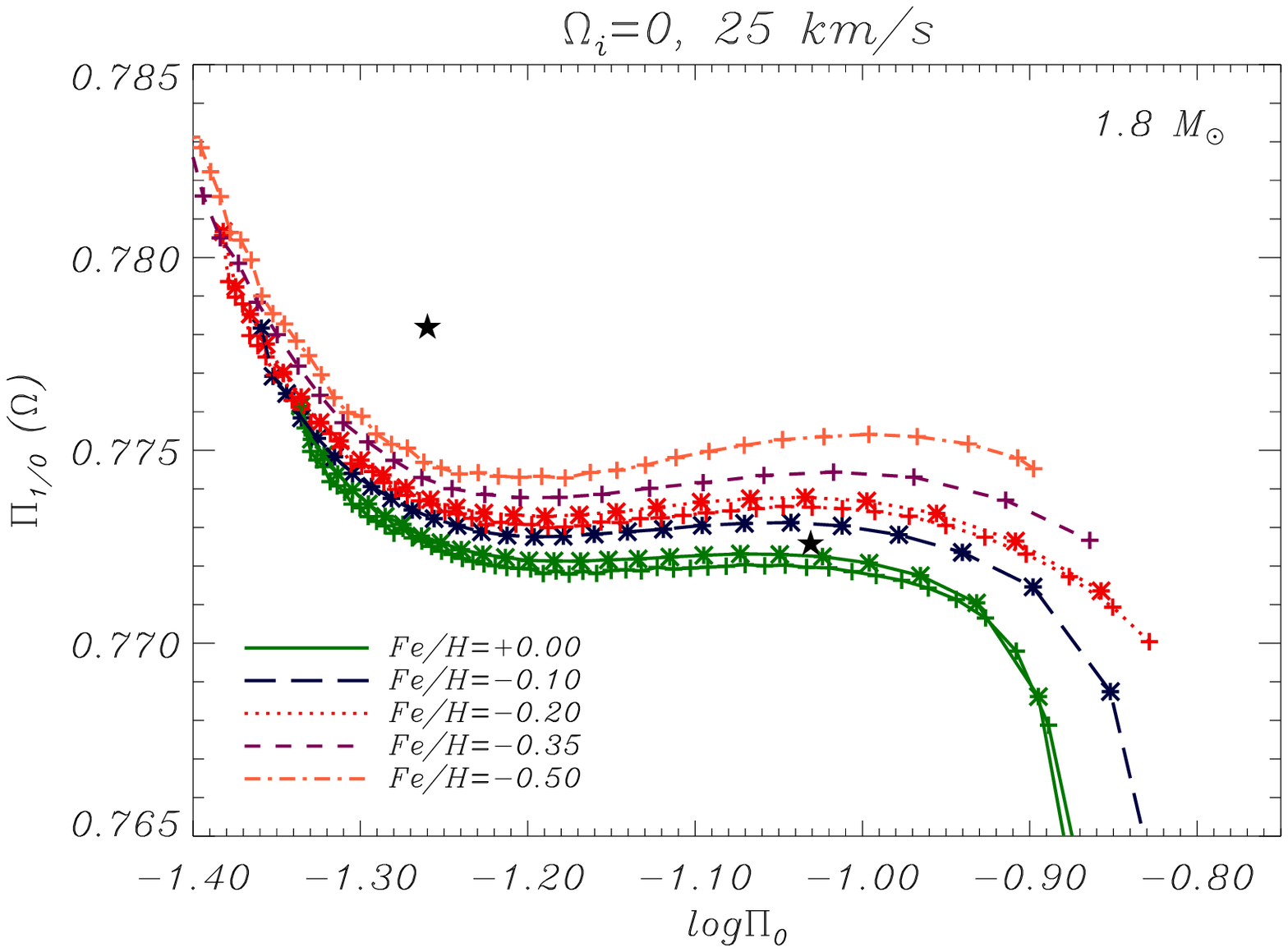}\hspace{-0.40cm} 
   \includegraphics[width=9cm]{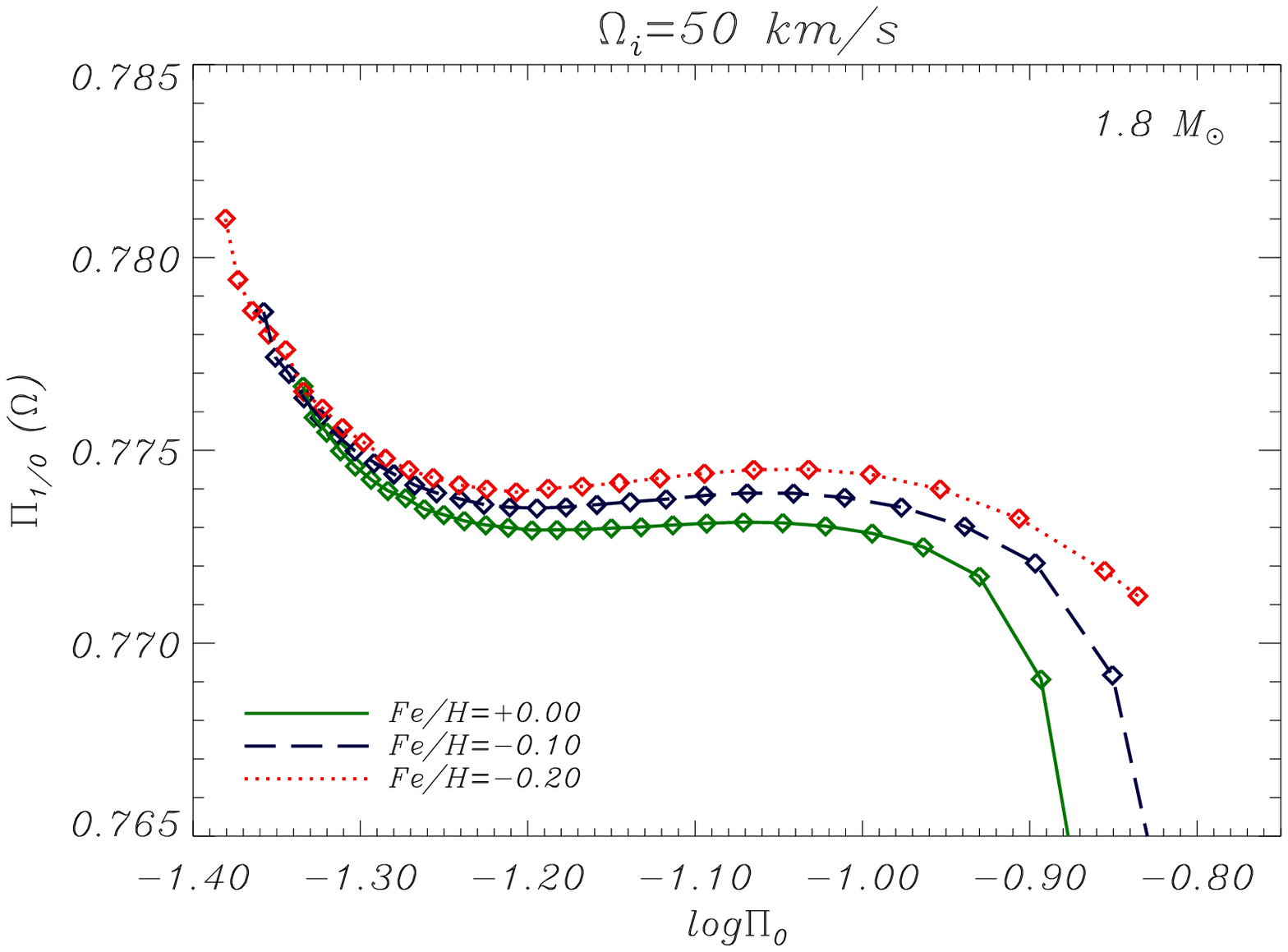}\hspace{-0.40cm}
   \includegraphics[width=9cm]{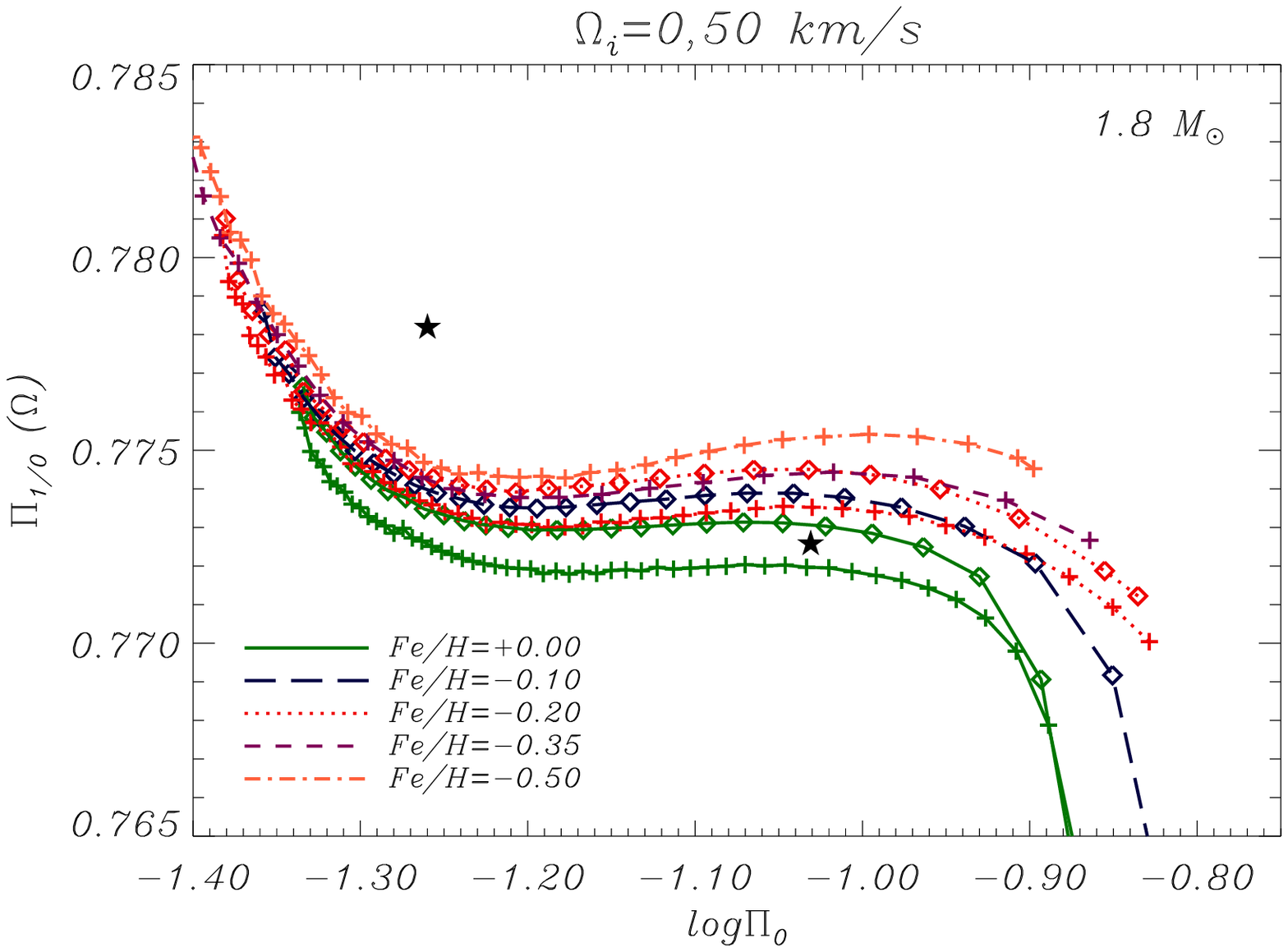}           
   \caption{Theoretical PD including rotation effects (RPD). The $\ratiorot$ period 
            ratios have been
            computed in the manner described in Sect.~\ref{sec:models} for a set
	    of evolutionary $1.8,\msol$ tracks obtained for different
	    metallicities. Tracks for three initial rotational velocities are 
	    considered: 25 and $50\,\kms$ (from top to bottom). Left panels
	    show only rotating models. Right panels show the comparison between
	    rotating and non-rotating tracks (classic PD). For convenience, the 
	    following symbols are used: crosses, representing non-rotating models;
	    asterisks, representing models evolved with $\Oi=25\,\kms$;
	    diamonds and those evolved with $\Oi=50\,\kms$. As in 
	    Fig.\ref{fig:classic_PD}, filled
	    star symbols represent the observed values for SX~Phe and RV~Ari.
	    (For clarity, colours are used in the \emph{on-line} version of the paper).}
   \label{fig:rot_PD}
 \end{center} 
\end{figure*}

\section{The $\ratiorot$ period ratios and the metallicity determinations
\label{sec:impact_Z}}

In order to analyse the impact of considering $\ratiorot$ rather than $\ratio$ period
ratios on metallicity determinations, these quantities, obtained from models 
taking into account rotation $\ratiorot$, are compared with those
$\ratio$. To do so, several evolutionary tracks have been computed as 
described in Sect.~\ref{sec:models} for six different metallicities:
$\mathrm{[Fe/H]}=0,-0.1,-0.2,-0.35,-0.50$ and $-1.00\,\mathrm{dex}$;
and three different initial rotational velocities $\Oi=25, 50$ and $100\,\kms$.
The case of rotational velocities between $50$ and $100\,\kms$ 
is purely illustrative, since there are no HADS known in that range.
The discussion is thus mainly focused on models with rotational velocities
up to $50,\kms$.
 
During the evolution, the rotational velocity of models decrease up to 
$0.75\,\Oi$ at TAMS, due to the global conservation of the total angular
momentum (see Sect.~\ref{sec:models} for more details). The mass of models is fixed 
to $1.8,\msol$ which typically corresponds to a \ds\ star.

Adiabatic oscillations are then computed from these rotating models, obtaining
thus the corresponding $\ratiorot$ period ratios. In Fig.~\ref{fig:rot_PD}
such \emph{rotational PD} (hereafter RPD) are displayed, from top to bottom, 
for tracks computed from $\Oi=25$ to $100\,\kms$ respectively. It can be noticed
that $\ratiorot$ period ratios increase for increasing rotational velocities
(left panels). Such effect is similar to decreasing the metallicity in classic
PD. As explained in the previous section, this can be understood in terms
of variations of the density distribution in stellar model interiors
due to rotation effects (Su\'arez et al., work in preparation).
In addition, the shift to larger period ratios is dependent on the metallicity.
In particular, the higher is the rotational velocity the closer are the
models of different metallicity in RPD (left panels). This means 
that the effect of rotation on period ratios is systematically larger for
increasing metallicity values (up to the solar value, in the present study).

In Fig.~\ref{fig:rot_PD} $\ratiorot$ (right panels) are displayed
together with classic ones $\ratio$. A first quantitative 
comparison is performed, in terms differences of period ratios 
$$\delta\ratio(\Omega,\mathrm{[Fe/H]})=\Big[\ratiorot-\ratio\Big]_{\mathrm{[Fe/H]}}$$
for a given metallicity. For the lowest initial rotational velocity 
considered, $\Oi=25\,\kms$, $\delta\ratio$ reach up $2-3\,10^{-3}$, and for
$\Oi=50\,\kms$, differences increase up to $6-8\,10^{-3}$

, and finally,
for $\Oi=100\,\kms$, they can be of the order of $10^{-2}$.
The main effect on PD (right panels) is to shift and compress
tracks of same metallicity (and mass) toward higher period ratios with
respect to classic tracks. 

The previous differences in period ratios can also be analysed
in terms of metallicity. For shortness, tracks
will be specified, from now on, with the subscript corresponding to the
rotational velocity considered. For instance, the track computed
with $\Oi=25\,\kms$ and $\mathrm{[Fe/H]}=-0.1$ will be 
called as $[-0.1]_{25}$.
\begin{figure}
   \includegraphics[width=8.8cm]{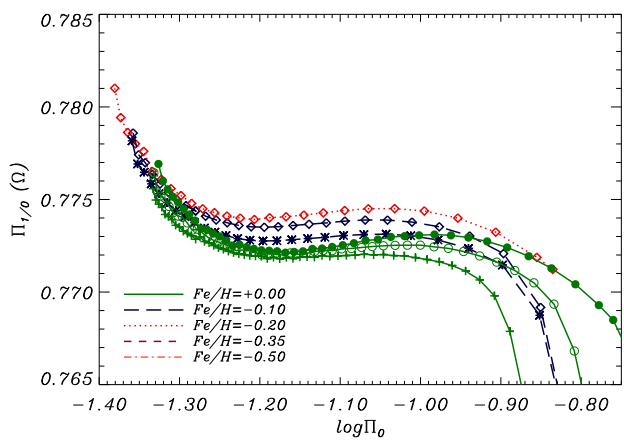}
   \caption{RPD ($\po$ in $\mathrm{d}$) illustrating the effect of
            considering different masses. Symbols have the same
	    meaning of those from Fig.~\ref{fig:rot_PD}, except for
	    empty and filled circles which correspond $1.9$ and $2.00\,\msol$
	    non-rotating models and solar metallicity respectively.
	    (For clarity, colours are used in the \emph{on-line} 
	    version of the paper).}
   \label{fig:rot_PD_M}
\end{figure}
As it will be shown, such analysis leads to a \emph{confusing} 
scenario: For $\Oi=25\,\kms$ (top, right) solar metallicity models tracks are 
similar to classic ones. However, when decreasing [Fe/H], rotating and
non-rotating tracks are located quite close. In particular, rotating 
$\mathrm{[-0.10]}_{25}$ tracks may be confused with $[-0.20]_{0}$ ones, and 
$\mathrm{[-0.20]}_{25}$ may be confused with non-rotating 
$\mathrm{[-0.35]}_{0}$ ones. The analysis of the middle right 
panel reveals that $\mathrm{[0.00]}_{50}$ tracks are located closely
to $\mathrm{[-0.20]}_{0}$ ones. Similarly, $\mathrm{[-0.10]}_{50}$ tracks
may be confused with  $\mathrm{[-0.20,-0.35]}_{0}$ ones, and finally,
$\mathrm{[-0.20]}_{50}$ tracks are close to $\mathrm{[-0.50]}_{0}$ ones.
As can be seen, the mix-up is critic for pop.~I stars
when considering $1.8\,\msol$ rotating models evolved with $\Oi=25, 50\,\kms$. 

Up to this point, all the discussion has been based on the results
obtained only for $1.8\,\msol$ models. When other masses and metallicites
are taken into account, the \emph{confusion} in metallicities and
rotational velocities significantly increases. 
In order to illustrate this, two solar metallicity tracks for
higher mass ($1.9$ and $2.00\,\msol$) non-rotating models 
are displayed in the RPD of Fig.~\ref{fig:rot_PD_M} where they 
are compared with other rotating non-solar tracks. As can be 
seen, the effect of increasing the mass of the models goes in the same
direction as increasing the rotational velocity, that is, it
increases the period ratios. It is worth noting that such shift toward 
larger period ratios systematically occurs for any \emph{rotating track}. 
Therefore the previous discussion based on $1.8\,\msol$ models 
is equivalent whatever the mass and/or rotational velocity considered.
Such behaviour may extend the \emph{confusing scenario} to
Pop.~II stars. Considering $1.8\,\msol$ models, it would be necessary
to consider initial rotational velocities up to $100\,\kms$. However,
this lower limit rapidly decrease when increasing the mass of the models.
For instance, $2\,\msol$ and $\Oi=50\,\kms$ tracks may be
misinterpreted with $\mathrm{[-1.00]}_{0}$ ones.

The construction of complete RPD for a wide range of metallicities, 
initial rotational velocities and masses becomes necessary for the 
exhaustive analysis of double-mode pulsators (work currently in progress), 
however this exceeds the scope of this paper.

\section{Conclusions\label{sec:conclusions}}

   The impact of taking into account the effect of rotation on Petersen Diagrams
   has been examined here, focusing on main sequence double-mode pulsators.
   Detailed seismic models have been computed considering rotation effects on
   both equilibrium models and on adiabatic oscillation frequencies. For $1.8\,\msol$
   stellar models, period ratios have been calculated for different rotational 
   velocities (Rotational Petersen Diagrams) and metallicities, and then compared 
   with classic non-rotating ones (PD). 
   
   The analysis of these RPD reveals that the difference in period ratios
   increases with the rotational velocity for a given metallicity.
   It remains around $10^{-3}$ for rotational velocities up to $50\,\kms$ and
   it can reach up $10^{-2}$ for rotational velocities close to $100\,\kms$.
   Such difference have been found enough to produce a significant 
   confusing scenario when analysing RPD in terms of metallicity variations.
   In particular, for $1.8\,\msol$ stellar models, differences in metallicity
   up to $\delta\mathrm{[Fe/H]}\sim0.30\,\mathrm{dex}$ can be found
   when considering models evolved with initial rotational velocities
   of $50\,\kms$. 
   Furthermore, such confusion may still increase when including other
   stellar masses, rotational velocities and metallicities, as well as
   other physical parametrisation (work in progress).
   
   The results presented here should thus be taken into account when
   analysing double-mode pulsators with Petersen Diagrams, in particular
   when accurate metallicity and/or mass determinations are required.
   
   A work on the detailed analysis of period ratios in 
   presence of near degeneracy is currently in preparation.
   Such work would provide new constraints on the modelling
   of HADS.


\acknowledgements{This study would not have been possible without the financial support from 
                  the European Marie Curie action MERG-CT-2004-513610. 
		  As well, this project was also partially financed
		  by the Spanish "Consejer\'{\i}a de Innovaci\'on, Ciencia y Empresa" from the
		  "Junta de Andaluc\'{\i}a" local government, and by
		  the Spanish Plan Nacional del Espacio under project 
		  ESP2004-03855-C03-01. JCS greatfully thank J. Christensen-Dalsgaard
		  and W. Dziembowski for their interesting and fruitful discussions
		  about this work.}
\bibliography{/home/jcsuarez/Boulot/Latex/Util/References/ref-generale}
\bibliographystyle{aa}

 \end{document}